\documentstyle[12pt]{ws-art}
\begin{document}
\title{Production of $K^-$-mesons in proton-proton and
proton-nucleus interactions at various energies}
\author{S.V. Efremov\\
{\it Bonner Nuclear Laboratory, Rice University, P.O. Box 1892,} \\
{\it Houston, TX 77251-1892, USA}
\and E.Ya. Paryev\\
{\it Institute for Nuclear Research, Russian Academy of Sciences,}\\
{\it Moscow 117312, Russia}}
\maketitle
\begin{abstract}The experimental data on the production of $K^-$-mesons in
pp-collisions are analyzed and a method of the unified description
of these data in a broad energy range for primary protons is proposed.
The $K^-$-mesons production in pA-collisions is considered.
The simple formulas for inclusive cross sections of the
$K^-$  production in these collisions are given.
The results of the calculations by these formulas are compared with
the available experimental data.
\end{abstract}

\section*{Introduction}

Experimental investigations of the production of $K^-$-mesons
in the nucleon-nucleon and nucleon-nucleus interactions have
been carried out for a long time (see, for example, Refs. [1 -- 14]).
It has been found that the inclusive
invariant cross section for the production of $K^-$-mesons in the
pp-collisions rapidly acquires scaling character as a function of
the transverse momentum, the radial scaling variable and the invariant
energy [15, 16] as the energy increases.
In a number of papers semiphenomenological
analytical expressions have been suggested, which describe
the experimental data on the $K^-$-mesons production in the
pp-collisions [16, 17] and pA-collisions
[18] at high energies. The direct experimental data
on the total and differential cross sections for the $K^-$-mesons
production in the pp-collisions at low energies are practically
absent. In the papers [19 -- 21] the expressions for the inclusive
cross sections for the $K^-$-mesons production in these collisions
near the threshold are given, which are based on the simple
phenomenological models. However, in view of the development
of high-current proton accelerators of a new generation -- kaon factories
 -- a need has arisen for a more detailed systematic description of
the experimental information on the inclusive cross sections for the production
of kaons in the nucleon-nucleon and nucleon-nucleus collisions
in a broad energy range for primary protons. Such a description
in the case of the $K^+$-mesons production in the pp- and
pA-interactions has been considered in [22]. In this work an attempt
of its generalization on the case of the $K^-$-mesons production
in these interactions has been undertaken. Apart from the purely
applied significance the obtained results may also serve
the interpretation of the experiments [10, 23, 24] on the production of
$K^-$-mesons in the nucleus-nucleus collisions at various
energies.

\section{The Total and Partial Cross Sections and Mean Multiplicities
for the Production of $K^-$-Mesons in the pp Collisions}

The experimental data on the production of $K^-$-mesons in
the pp-collisions are available
mainly in the region of high energies (at kinetic energy of a primary
proton in the lab system
$\epsilon_0>10$ GeV) in the form of mean multiplicities and inclusive
spectra, while the data in the low-energy region (at $\epsilon_0\leq 10$ GeV)
are rather scarce. Let us consider at first the low-energy region.
The experimental data on the total cross section for the $K^-$ production
in the pp-collisions, $\sigma_{pp\rightarrow K^-X}$ are practically
absent at present. The authors know only one value of the cross section
$\sigma_{pp\rightarrow K^-X}$ at $\epsilon_0=2.85$ GeV, which is equal to
$4\pm 2$ $\mu$b [1]. This value is presented in Fig.~1. For the evaluation
of $\sigma_{pp\rightarrow K^-X}$ at $\epsilon_0\sim3\div7$ GeV let us use
the available data on the partial cross section for the production
of $K^-$- and $\bar{K}^0$-mesons
in the pp-interactions. The reaction of the $K^-$ production in these
interactions, which has the lowest threshold ($\epsilon_{thr}=2.49$ GeV)
is the following reaction with four particles in the final state
\begin{equation}p+p\,\rightarrow\,p+p+K^{+}+K^{-}.\end{equation}
At present there is only one experimental value of the total cross section
of this reaction $\sigma_{pp\rightarrow ppK^+K^-}$ at $\epsilon_0=5.7$ GeV,
equal to $34\pm12$ $\mu$b, which is given in [19].\footnote{Note that
the experimental investigation of the reactions of the production of
$K\bar{K}$-pairs in the pp-collisions in the energy range under
consideration is planned at the accelerator
CELSIUS (Uppsala, Sweden) in the near future [25].}
Therein this cross section at $\epsilon_0\leq7$ GeV
was calculated in the framework of
one-kaon exchange model. The results of the calculation
are represented in Fig.~1
by a dash-dotted curve. Beginning from the energies $\epsilon_0\sim5\div6$ GeV
an appreciable contribution into the $\sigma_{pp\rightarrow K^-X}$ is given
by the reactions with one and two additional $\pi$-mesons in the final state
along with the reaction (1) [2, 26]. The existing experimental data on the
total cross sections of these reactions are given in Table 1.\newpage
\begin{center}
\centerline{\bf Table 1. Partial cross sections of the reaction
 $pp\rightarrow pN
K^{-}K^{0}\nu\pi$,}
\centerline{\bf $\nu=1,2$ [2, 26]}
\centerline{ }
\begin{tabular}{|l|c|c|c|}\hline
Reaction  & $\epsilon_0,$ GeV & $\epsilon_{thr}$, GeV & $\sigma$,
$\mu$b\\ \hline
$p+p\,\rightarrow\,p+p+K^-+K^0+\pi^++\pi^0$ & 6.05 & 3.38 & 2.6$\pm$2.5\\
\hline
$p+p\,\rightarrow\,p+p+K^-+K^0+\pi^+$      & 6.99 & 2.94 & 13.8$\pm$5.2\\
\hline
$p+p\,\rightarrow\,p+p+K^-+K^0+\pi^++\pi^0$& 6.99 & 3.38 & 5.9$\pm$3.4\\ \hline
$p+p\,\rightarrow\,p+n+K^-+K^0+2\pi^+$     & 6.99 & 3.41 & 3.9$\pm$2.8\\ \hline
\end{tabular}\\
\end{center}
\centerline{ }
{}From
these data and from the data on $\sigma_{pp\rightarrow ppK^+K^-}$ presented
in Fig.~1 it follows particularly that at $\epsilon_0=7$ GeV
$\sigma_{pp\rightarrow K^-X}\approx 0.08$ mb. Note that at this energy
the total cross section for the $K^+$-mesons production
$\sigma_{pp\rightarrow K^+X}$ constitutes a value of 0.9 mb [22].
To obtain an experimental estimate of the value of $\sigma_{pp\rightarrow
K^-X}$
at $\epsilon_0\sim4\div5$ GeV let us suppose that
$\sigma_{pp\rightarrow K^-X}=\sigma_{pp\rightarrow \bar{K}^0X}$ [21].
\footnote{It is easily seen that according to the hypothesis of the
invariance of the strong interactions relative to the rotations in
the isotopic space, the total cross sections for the production of $K^-$-
and $\bar{K}^0$-mesons in the pn-collisions are equal to each other~[27].}
Using the sum of the partial cross sections of the following channels [26]
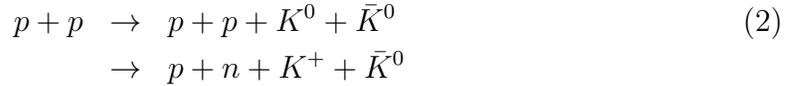
\footnote{One can neglect the contribution of the reactions
with additional pions in the final state in $\sigma_{pp\rightarrow \bar{K}^0X}$
at $\epsilon_0\sim4\div5$ GeV [2, 26].}
\begin{eqnarray}
p+p&\rightarrow&p+p+K^{0}+\bar{K}^{0}\\
&\rightarrow&p+n+K^++\bar{K}^{0}\nonumber
\end{eqnarray}
as a value of $\sigma_{pp\rightarrow \bar{K}^0X}$ in this energy range
we obtain that  $\sigma_{pp\rightarrow K^-X}=15.0\pm3.2$
$\mu$b at $\epsilon_0=4.1$ GeV, while
$\sigma_{pp\rightarrow K^-X}=25.8\pm11.1$
$\mu$b at $\epsilon_0=4.6$ GeV. These values are also presented in Fig.~1.
Therein the dashed curve
denotes the results of the calculation of
$\sigma_{NN\rightarrow NNK\bar{K}}$ by the formula
suggested in [21] on the basis  of the experimental
data analysis [26] on the total cross sections of the reactions
(2) at $4.1\leq\epsilon_0\leq9.1$ GeV.
\begin{equation}
\sigma_{NN\rightarrow NNK\bar{K}}(s)=\frac{1}{40}
\stackrel{\ast}{p}_{max}\, mb,\end{equation}
This formula was used in [21] as a total
cross section for the production of $K^-$-mesons in the nucleon-nucleon
collisions with four particles in the final state. Here
\begin{equation}
\stackrel{\ast}{p}_{max}=(2\sqrt{s})^{-1}(s-4(m_N+m_K)^2)^{1/2}
(s-4m_N^2)^{1/2}\end{equation}
is the maximum momentum (in GeV/c) of $\bar{K}$-meson in the NN -- c.m.s.
for the reaction $NN\rightarrow NNK\bar{K}$; $m_N$ and $m_K$ are the rest
masses of nucleon and kaon; and $s$ is the invariant energy squared .
It is seen that calculations by the one-kaon exchange model and by
the formula (3), which, according to the above present,
$\sigma_{pp\rightarrow K^-X}$ at $\epsilon_0\leq5$ GeV, do not agree
with each other and with the experiment near the
threshold (the arrow in Fig.~1). Thus, it is not
correct to use these calculations for the above cross section in the
energy range of $\epsilon_0>5$ GeV.

For the description of $\sigma_{pp\rightarrow K^-X}$ at $\epsilon_0>5$ GeV
let us invoke some additional information, namely, the data on the mean
multiplicity of $K^-$-mesons in the pp-collisions $<n_{K^-}>$. The
experimental information on $<n_{K^-}>$ was obtained in [3] and listed in
Table 2.
\centerline{ }
\begin{center}
\centerline{\bf Table 2. Mean multiplicities of $K^{-}$-mesons in the pp-
collisions}
\centerline{ }
\begin{tabular}{|c|c|}
\hline
s, GeV$^2$     & $<n_{K^-}>$\\ \hline
25.3           & 0.008  \\ \hline
37.8           & 0.036  \\ \hline
46.8           & 0.033  \\ \hline
67.2           & 0.07   \\ \hline
81.0           & 0.08   \\ \hline
100            & 0.11   \\ \hline
133            & 0.13   \\ \hline
485            & 0.24   \\ \hline
960            & 0.29   \\ \hline
2025           & 0.34   \\ \hline
2810           & 0.37   \\ \hline
\end{tabular}
\end{center}
\centerline{ }
Knowing the mean multiplicity $<n_{K^-}>$, it is easy to find
the total cross section for the $K^-$-meson production
in the pp-collisions, since
\begin{equation}
\sigma_{pp\rightarrow K^{-}X}=\sigma_{pp}^{inel}<n_{K^-}>.\end{equation}
Here $\sigma_{pp}^{inel}$ is the inelastic pp-interaction cross section.
The results of the calculation of the cross section
$\sigma_{pp\rightarrow K^-X}$ by Eq. (5)
is presented in Fig.~2.\footnote{The data on $\sigma_{pp\rightarrow K^-X}$
at low energy which are shown in Fig.~1 are also presented here.}.It was
supposed in this calculation that
$\sigma_{pp}^{inel}=30$ mb at $s<100$ GeV$^2$ and
$\sigma_{pp}^{inel}=32$ mb at $s\geq 100$ GeV$^2$ [3, 28] and it was also
considered that errors in $<n_{K^-}>$ were equal to 25\% at
$67.2\leq s\leq133$ GeV$^2$ and 15\% at the other values of $s$ [3].
We shall seek the parameterization of these data in a  form, such that it
would become the expression given by the phase space model [29]:
\begin{equation}\sigma_{pp\rightarrow K^-X}\sim(s-s_{min}^-)^{3.5},
\end{equation}
as $s\rightarrow s_{min}^-$ ($\sqrt{s_{min}^-}=2(m_p+m_K)=2.8639$ GeV),
while at $s\rightarrow \infty$ it would satisfy the following
relation [3, 30, 31]
\footnote{It is necessary to note that this relationship is actually
true beginning from $s\sim100$ GeV$^2$. It reflects the fact that the mean
multiplicities of the particles of various type ($\pi^{\pm}$, $K^{\pm}$,
 $\bar{p}$) produced in the pp-collisions may be fitted by one universal
curve, beginning from the Serpukhov's energies, if one simply redefines
the gauge of the energy and changes the normalization [30, 31].}
\begin{equation}\sigma_{pp\rightarrow K^-X}(s)=
\sigma_{pp\rightarrow K^+X}(s/s_0), \end{equation}
where $s_0=2.6\pm0.5$ GeV$^2$ [30]. Taking into account that [22]
$$\sigma_{pp\rightarrow K^+X}(s)=\sigma_1^+F_1(s/s_{min}^+)+\sigma_2^+
F_2(s/s_{min}^+),$$
$$F_1(x)=(1+1/\sqrt{x})\ln{(x)}-4(1-1/\sqrt{x}),$$
\begin{equation}F_2(x)=1-(1/\sqrt{x})(1+\ln{(\sqrt{x})});
\end{equation}
\begin{eqnarray}F_1(x)\approx(x-1)^3/3,&F_2(x)\approx(x-1)^2/8&
for\,\,x\rightarrow 1;\nonumber\\
F_1\approx\ln{(s)},&F_2\approx1 & for\,\,s\rightarrow\infty;\nonumber\\
\sigma_1^+=2.8\pm0.8\,\,mb,&
\sigma_2^+=9.7\pm1.5\,\,mb,&\sqrt{s_{min}^+}=
m_p+m_{\Lambda^0}+m_K=2.5476\,\,GeV,\nonumber\end{eqnarray}
and according to (6), (7) we choose the dependence of
$\sigma_{pp\rightarrow K^-X}(s)$ to be the following form
\begin{equation}
\sigma_{pp\rightarrow K^-X}(s)=\left(1-\frac{s_{min}^{-}}{s}
\right)^k\left[\sigma_{1}^{-}F_1\left(\frac{s}{s_{min}^{-}}\right)
+\sigma_{2}^{-}F_{2}\left(\frac{s}{s_{min}^{-}}\right)\right]
+\sigma_3^-F_3\left(\frac{s}{s_{min}^{-}}\right),
\end{equation}
\begin{eqnarray}F_3(x)=\left(\frac{x-1}{x^2}\right)^{3.5},&
\sigma_1^-=\sigma_1^{+},&\sigma_2^{-}=\sigma_2^{+}+
\sigma_1^{+}\ln\left(\frac{s_{min}^{-}}{s_{min}^{+}s_0}\right)=
7.7\pm1.5\,\,mb.\nonumber\end{eqnarray}

	The coefficients $k$ and $\sigma_3^-$ in (9), which have been found
using the data presented in Fig.~2, are  $k=3$ and
$\sigma_3^-=3.90\pm0.85$ mb.
The calculation of the cross section $\sigma_{pp\rightarrow K^-X}$ according to
(9) is represented in Figs.~1,~2 by the solid curve. It is seen that the
approximation (9) well describes the existing\footnote{
In the sense which was outlined above} experimental data in a broad range of
the primary proton energies. In this case, particularly,
$\sigma_{pp\rightarrow K^-X}=85$ $\mu$b at $\epsilon_0=7$ GeV.
\section{The Invariant Inclusive Cross Section for the
Production of $K^-$-Mesons in the pp Interactions}

There are experimental data on the differential cross sections
for the production of $K^-$-mesons in the pp-collisions only at
$\epsilon_0>10$ GeV. For the description of these data it is
natural to choose the parameterization of the invariant inclusive
cross section $E_{K^-}d\sigma_{pp\rightarrow K^-X}/d{\bf p}_{K^-}$ in
the form used by us in [22] for the analogous
cross section for the production of $K^+$-mesons, namely:
\begin{equation}
E_{K^-}\frac{d\sigma_{pp\rightarrow K^-X}}{d{\bf p}_{K^-}}
=B(s)(1-x_R)^{n(s)}
\exp{\left[b-\sqrt{b^2+c^2(s)p_{\perp}^2}\right]},\end{equation}
\begin{equation}
x_R=\stackrel{\ast}{E}_{K^-}/(\stackrel{\ast}
{E}_{K^-})_{max}=\frac{2(E_0+m_p)E_{K^-}-
2p_0p_{K^-}\cos{\vartheta}}{s-(2m_p+m_K)^2+m_K^2},
\end{equation}
where $p_{\perp}$ is the transverse momentum of a $K^-$-meson, $\stackrel{\ast}
{E}_{K^-}$ and $(\stackrel{\ast}{E}_{K^-})_{max}$ are the total energy
of the $K^-$-meson in the centre-of-mass system and its maximum possible value,
$E_0$, $E_{K^-}$ and ${\bf p}_0$, ${\bf p}_{K^-}$ are the energies and
momenta in the lab system of a primary proton and kaon, respectively, and
$\cos{\vartheta}={\bf p}_0{\bf p}_{K^-}/p_0p_{K^-}$. From the comparison with
the experimental data at the initial momenta $p_0=12.5$ GeV/c [4] and $24$
GeV/c [5]; and from the  threshold behaviour given by
the phase space volume model
and scaling behaviour [6, 15, 16] the following
dependences of the parameters in Eq (10) on s were found:
$$B(s)=\left(200+2600\exp{\left[-\frac{(x-1)^2}{0.2}\right]}-340
\exp{\left[-\frac{(x-0.6)^2}{0.03}\right]}\right)
\,\frac{GeV\,\mu b}{(GeV/c)^{3}\,sr},$$
\begin{eqnarray}
n(s)=2+3.8x,&b^2=3,&c^2(s)=23x,\end{eqnarray}
$$x=w/(0.3+w+0.63e^{-2w}),\,\,
w=(\sqrt{s/s_{min}^-}-1)^2.$$

	The comparison of the results of our calculations by Eqs. (10) -- (12)
for the double differential cross sections for the production of $K^-$-mesons
in the pp-collisions with the experimental data [4 -- 8] for
$12.5\leq p_0\leq1500$ GeV/c is given in Figs.~3~--~8. It can be seen that
for $p_{\perp}\leq1.5$ GeV/c there is a reasonable agreement of
our calculations with the experiment in a broad energy range of a primary
protons. It allows one to hope for the applicability of expressions
(10) -- (12) at any high energy. Moreover, since the
parameters in the approximation (10) were obtained particularly from its
threshold behaviour one can hope
for the applicability of these expressions at low energies, taking into
account the smooth dependence of
$E_{K^-}d\sigma_{pp\rightarrow K^-X}/d{\bf p}_{K^-}$ on $s$. Naturally,
in order to check directly the truth of the last assumption it is necessary
to have the appropriate experimental data. The experimental information
[9, 10] on the production of $K^-$-mesons in the collisions of protons
with light nuclei (see the next section) may serve as another additional
criterion for testing the applicability of formulas (10) -- (12)
at low energies. And finally, another important criterion for verification
of the results obtained is the comparison of the total $K^-$-meson production
cross sections found from the data on the multiplicity, partial
cross sections for the production of $\bar{K}^0$-mesons and from the
double differential distributions. Therefore in Figs.~1 and 2
the dashed curve with two points represents
our calculation of $\sigma_{pp\rightarrow K^-X}$ by the
numerical integration of the expressions (10) -- (12).
It is evident that the cross section values thus found are close to those
calculated by Eq. (9) and for $s>20$ GeV$^2$ they even slightly
underestimate them (by the value of the order of 30\%).

Thus, the analysis carried out shows that formulas (9) for
$\sigma_{pp\rightarrow K^-X}$ and (10) -- (12) for
$E_{K^-}d\sigma_{pp\rightarrow K^-X}/{\bf p}_{K^-}$ describe the whole
set of the experimental data on the production of $K^-$-mesons
in the pp-collisions with an accuracy  of 30 -- 40\%.

\section{Production of $K^-$-Mesons in Proton-Nucleus Collisions }

We consider at first the production of $K^-$-mesons in the collisions of
protons with light nuclei. In such collisions the largest fraction of kaons is
made up by those which are produced in direct proton-nucleon collisions,
i.e. the production of kaons happens in the single event of the inelastic
interaction of a primary proton with one of the intranuclear nucleons.
The inclusive cross section
for this process may approximately be represented as follows [22]:
\begin{equation}
E_{K^-}\frac{d\sigma_{pA\rightarrow K^-X}}
{d{\bf p}_{K^-}}=I_V(A)
E_{K^-}\frac{d\sigma_{pp\rightarrow K^-X}}{d{\bf p}_{K^-}},\end{equation}
\begin{equation}
I_V(A)=A\int\rho({\bf r})\,
d{\bf r}\,\exp{\left[-\mu(p_0)\int\limits_{-\infty}^{0}
\rho({\bf r}+x'{\bf\Omega_0})\,dx'-\mu(p_{K^-})
\int\limits_{0}^{\infty}\rho({\bf r}+x'{\bf\Omega})\,dx'
\right]};\end{equation}
$$\mu(p_0)=\sigma_{pp}^{inel}(p_0)Z+
\sigma_{pn}^{inel}(p_0)N,$$
\begin{equation}
\mu(p_{K^-})=\sigma_{K^-p}^{inel}(p_{K^-})Z+
\sigma_{K^-n}^{inel}(p_{K^-})N,
\end{equation}
where $\sigma_{K^-p}^{inel}$ ($\sigma_{K^-n}^{inel}$) is the
inelastic free-particle cross section
for the $K^-$-p ($K^-$-n) interaction, $\rho({\bf r})$ is the average density
of nuclear nucleons
at a point ${\bf r}$, normalized to unity, $Z(N=A-Z)$ is the number
of protons (neutrons) in a nucleus, and ${\bf\Omega}_0={\bf p}_0/p_0$,
${\bf\Omega}={\bf p}_{K^-}/p_{K^-}$. Here, the effect of the Fermi
motion of nuclear nucleons on the kaon production in the collision
of a primary proton with an intranuclear nucleon is neglected,
since in the given region of initial energies, emission angles, and
momenta of $K^-$-mesons, as shown by our calculations, its inclusion is
unimportant. In addition, the $K^-$-meson production
cross sections in the pp and pn-interactions are supposed to be the
same [21, 27]. The inelastic cross sections of the proton-nucleon
and kaon-nucleon interactions are taken to approximately describe the
possibility of the $K^-$-meson production by a primary proton
after proton's "soft" elastic rescatterings on the intranuclear nucleons
and the chance of the "survival" of the kaon in a given interval of angle and
momentum  after kaon's drastically anisotropic [32] elastic
rescatterings on the nucleons of the nucleus. Thus, in the framework
of the single interaction model the knowledge of the cross section for
the production of kaons in the
nucleon-nucleon collisions allows one to calculate the cross section
for kaons production in the nucleon-nucleus collisions according to the
formulas (13) -- (15). And, conversely, the experimental information
about the production of kaons on light nuclei by the protons may be
useful for the additional check-up of kaon production models in pp-collisions.

In the case where ${\bf\Omega}\approx{\bf\Omega}_0$, and for the
Gaussian distribution of the density of nuclear nucleons
($\rho({\bf r})=(b_0/\pi)^{3/2}\exp{(-b_0r^2)}$) we approximately
have [22]
\begin{equation}I_V(A)=A(1-e^{-x_G})/x_G,
\,\,\,x_G=\left[\mu(p_0)+\mu(p_{K^-})\right]b_0
/2\pi.\end{equation}
Note that in [22] the integral (14) has been calculated also
for the nucleus with the uniform nucleon density and for $\mu(p_K)=0$
it has been reduced to a more simple form.

In Fig.~9 we compare the results of our calculations by  Eqs.
(10) -- (16) (solid curve) with the experimental
data [9] for the
inclusive invariant cross section for the $K^-$-meson production
at an angle of 3.5$^0$ in the interaction of protons with
$p_0=10.1$ GeV/c with Be nuclei. In the same place the results of
calculations (dashed curve) are presented, where in (15) instead
of the inelastic cross sections for the kaon-nucleon interaction
$\sigma_{K^-N}^{inel}$ the total cross sections for this interaction
$\sigma_{K^-N}^{tot}$ were used.
The total cross sections for the inelastic proton-nucleon
interaction and parameter $b_0$ hereafter were taken to be equal to
30 mb and 0.24 fm$^{-2}$, respectively [22]. The total and elastic
($\sigma_{K^-N}^{el}$) cross sections for the $K^-$-meson-nucleon
interaction were borrowed from [33] and parametrized as follows
\begin{eqnarray}
\sigma_{K^-p}^{tot}(p_{K^-})=&22.6p_{K^-}^{-1.14},&0.245\leq{p_{K^-}}\leq0.7,
\nonumber\\
                   =&47.54p_{K^-}^{0.94},&0.7\leq{p_{K^-}}\leq1.1,\nonumber\\
                   =&69.87p_{K^-}^{-3.1},&1.1\leq{p_{K^-}}\leq1.3,\nonumber\\
       		   =&32.4p_{K^-}^{-0.17},&1.3\leq{p_{K^-}}\leq10,\nonumber\\
		   =&22,                 &10 \leq{p_{K^-}}\leq310,\nonumber\\
\sigma_{K^-p}^{el}(p_{K^-})=&10.58p_{K^-}^{-0.98},&0.03\leq{p_{K^-}}\leq0.7,\\
                   =&23p_{K^-}^{1.2},    &0.7 \leq{p_{K^-}}\leq1,\nonumber\\
  		   =&23p_{K^-}^{-2.6},   &1   \leq{p_{K^-}}\leq1.5,\nonumber\\
		   =&9.56p_{K^-}^{-0.44},&1.5 \leq{p_{K^-}}\leq20,\nonumber\\
		   =&2.56,     		 &20  \leq{p_{K^-}}\leq310;\nonumber
\end{eqnarray}
\begin{eqnarray}
\sigma_{K^-n}^{tot}(p_{K^-})=&36.7p_{K^-}^{0.75}  &0.63\leq{p_{K^-}}\leq1,
\nonumber\\
		   =&36.7p_{K^-}^{-0.726}&1   \leq{p_{K^-}}\leq1.95,\nonumber\\
		   =&23.23p_{K^-}^{-0.05}&1.95\leq{p_{K^-}}\leq20,\nonumber\\
		   =&20                  &20  \leq{p_{K^-}}\leq310,\\
\sigma_{K^-n}^{el}(p_{K^-}) =&29.2p_{K^-}^{3.59},
&0.612\leq{p_{K^-}}\leq0.9,\nonumber\\
		   =&17.1p_{K^-}^{-1.51},&0.9  \leq{p_{K^-}}\leq3,\nonumber\\
		   =&3.25,               &3    \leq{p_{K^-}}\leq310.\nonumber
\end{eqnarray}
Here the momenta are measured in GeV/c and the cross sections
are in mb. The inelastic cross sections for the $K^-$-meson-nucleon
interaction were calculated as a difference between the total
and elastic cross sections. It is seen that
the calculation is in reasonable agreement with the experiment,
although there is a small discrepancy in the softest part of the
spectrum, which seems to be caused by the rescattering of kaons
in the nuclear medium (which was taken into account only
approximately in deriving expression (13)), as well as by the
contribution to this part of the spectrum of various channels
(particularly of the channel $K^-N\,\rightarrow\,K^-N\nu\pi$, $\nu=$1, 2...)
of the reaction $K^-N\,\rightarrow\,K^-X$ [33].

In Fig.~10 we show the energy dependence of the inclusive invariant cross
section for the production of $K^-$-mesons with a momentum of 0.8 GeV/c
and at an angle of 24$^0$ by protons on the $^{12}$C nucleus
compared with the experimental data [10]. It is evident that at
$\epsilon_0<5$ GeV there is a satisfactory agreement of our calculations
with the experiment. The discrepancy
with the experiment at $\epsilon_0>5$ GeV as well as in the previous
case seems to be due to the increasing contribution of inelastic
$K^-N$-interactions to the cross section for $K^-$-mesons at higher energies
in the considered energy range.

In Figs.~11 and 12 we present  the comparison of our calculations
of the double differential cross sections for the
production of $K^-$-mesons by protons on Be nuclei with
the experimental data [11, 12]: in Fig.~11 -- by protons with momenta
18.8 (dashed curves) and 23.1 GeV/c (solid curves) at angles of 0 and 5.7$^0$;
and in Fig.~12 -- at $p_0=400$ GeV/c for $K^-$-mesons with $p_{\perp}=0$ (solid
curve) and 0.5 GeV/c (dashed curve). The absorption cross section of the
primary proton by these nuclei in the calculations was assumed to be equal
to 188 mb [22]. It is seen that there is a satisfactory agreement here
as well.

Thus the results obtained show that the simple analytical formulas
proposed (which are analogous to those obtained in [22] for $K^+$-mesons)
allow one to calculate satisfactorily the inclusive cross sections for
the production of $K^-$-mesons at small angles on light nuclei
by protons of various energies.

The calculation of the $K^-$-mesons production on heavy nuclei
requires the inclusion of some additional factors, particularly
of those which were outlined previously. However, as it follows
from Refs. [13, 34], the experimental data on the kaon production
cross section on heavy nuclei, at least in the region of the
energies $\epsilon_0=24\div400$ GeV for
$p_{\perp}\leq 1$ Gev/c, are approximately described by  the dependence
analogous to (13), where
$$I_V(A)=\left(\frac{A}{A_{Be}}\right)^{\alpha(x_F,p_{\perp})
}I_V(Be);$$
\begin{equation}\alpha(x_F,p_{\perp})=
\alpha_1(x_F)+\alpha_2(p_{\perp}),\end{equation}
$$\alpha_1(x_F)=0.74-0.55x_F+\frac{x_F}{\mid x_F
\mid}0.26x_F^2,\,\,x_F=\stackrel{\ast}{p}_{\parallel}/
(\stackrel{\ast}{p}_{\parallel})_{max},$$
$$\alpha_2(p_{\perp})=0.1p_{\perp}^2.$$
This fact is obviously illustrated by Figs.~13 and 14.
As one can see from these figures, the  experiment is well fitted by
the calculations.
\section*{Conclusion}
The analysis carried out in this work showed that simple
analytical formulas proposed  for the inclusive cross sections of the
production of $K^-$-mesons in the pp and pA-interactions satisfactorily
describe the experimental data in a broad range of energies of
primary protons. Therefore one can believe that the predictions for
the cross sections, being obtained by their use, will appear to be sufficiently
reliable and may be used in connection with the projects of the
accelerators of a new generation -- kaon factories.

We note, finally, that the maximum uncertainty in the data on the $K^-$-meson
production in the nucleon-nucleon collisions refers to the
energy region $\epsilon_0<10$ GeV. Therefore, for the better understanding
of the kaon production phenomenon in the nucleon-nucleon collisions
it is necessary to carry out the experimental
measurements of the $\sigma_{pp\rightarrow K^-X}$ and/or
$E_{K^-}d\sigma_{pp\rightarrow K^-X}/d{\bf p}_{K^-}$ in this energy range.

In conclusion the authors express the gratitude to M.V. Kazarnovsky
 for the discussion of some questions which have arisen in the
solution of this problem.

\newpage
\centerline{\bf Figure Captions}
\centerline{ }
{\bf Fig.1.} Dependence of the total $K^-$-mesons production cross section on
the
 energy in the pp-collisions. Calculations: the dot-dash line, the
 cross section $\sigma_{pp\rightarrow ppK^+K^-}$
calculated using the one-kaon exchange model [19]; the dashes, by (3);
the solid curve, by (9); two dots
 and a dash, by the numerical integration of expressions (10)--(12).
 The experimental data: see text.\\

{\bf Fig.2.} Dependence of the total $K^-$-mesons production cross section on
the
 invariant energy squared $s$ in the pp-collisions. The notations are
 the same as in Fig.1.\\

{\bf Fig.3.} Double differential cross sections for the production of
$K^-$-mesons,
$d^2\sigma/dpd\Omega$, in the $p+p\rightarrow K^-+X$ reaction at
$p_0=12.5$ GeV/c.\\
a) Dependence of the $d^2\sigma/dpd\Omega$ on the longitudinal momentum
$\stackrel{\ast}{p}_{\parallel}$ of a
$K^-$-meson in the centre-of-mass system at $p^2_{\perp}=0.4$ (GeV/c)$^2$;\\
b) dependence of the $d^2\sigma/dpd\Omega$ on the $p^2_{\perp}$ at
$p_{\parallel}=0.6$ GeV/c. The experimental data: [4].
The solid curves, calculations by (10)--(12).\\

{\bf Fig.4.}  Inclusive invariant cross sections for the production of
$K^-$-mesons
 in the pp-collisions at $p_0=24$ GeV/c as functions of
$\stackrel{\ast}{x} = \stackrel{\ast}{p}_{\parallel}/\stackrel{\ast}{p}_{max}$
at fixed values (indicated by numbers by the curves) of their transverse
momentum. The solid curves, the experimental data [5]. The
 dashed curves, calculations by (10)--(12).\\

{\bf Fig.5.} Inclusive invariant cross section for the production of
$K^-$-mesons
 in the pp-collisions at $p_0=24$ GeV/c as a function of their transverse
momentum $p_{\perp}$ at a fixed value of the variable $\stackrel{\ast}{x}$.
The notations are the same as in Fig.4.\\

{\bf Fig.6.} Inclusive invariant cross sections for the production of
$K^-$-mesons
in the pp-collisions at different values of the invariant energy
 squared s (indicated by numbers in the figure) as functions of their
       transverse momentum at fixed angles $\stackrel{\ast}{\vartheta}$
in the centre-of-mass system
 for given s (tg$\stackrel{\ast}{\vartheta} = 2\alpha/(s)^{1/2}$,
$\alpha=1.33$ GeV/c).
The experimental data: [6].
The solid and dashed curves, our calculations by (10)--(12) at $s=949$
and $s=47$ GeV$^2$, respectively.\\

{\bf Fig.7.} Inclusive invariant cross sections for the production of
$K^-$-mesons
in the pp-interactions at different collision energies $(s)^{1/2}$
       (indicated by numbers in the figure) as functions of
$x_F^`=2\stackrel{\ast}{p}_{\parallel}/(s)^{1/2}$ at
       fixed values of their transverse momentum $p_{\perp}$. The experimental
       data: [8]. The solid curves, calculations by (10)--(12)
at $(s)^{1/2}=53$ GeV.\\

{\bf Fig.8.} Inclusive invariant cross section for the production of
$K^-$-mesons
 in the pp-interactions at different collision energies $(s)^{1/2}$
       as a function of their transverse momentum $p_{\perp}$
at a fixed value of the variable $x^`_F$.
The notations are the same as in Fig.7.\\

{\bf Fig.9.}  Inclusive invariant cross section for the production of
$K^-$-mesons
at an agle of $3.5^0$ in the interaction of protons of momentum $10.1$ GeV/c
with Be nuclei as a function of the $K^-$-meson momentum.\\

{\bf Fig.10.}  Inclusive invariant cross section for the production of
$K^-$-mesons
  with momentum of $0.8$ GeV/c  at an agle of $24^0$ on $^{12}$C nuclei as a
 function of primary-proton energy. The solid and dashed curves
 denote the same as in Fig.9. The dot-dashed curve denotes the
 same as dashed curve, but it is supposed that the $K^-$-meson production
cross section in the pn-collisions is larger than one in
 the pp-collisions by a factor of two [20].\\

{\bf Fig.11.}  Double differential cross sections for the production of
$K^-$-mesons at angles of 0 and 5.7$^0$ in the interaction of protons
of momenta $18.8$ and $23.1$ GeV/c with Be nuclei as functions of the
$K^-$-meson momentum.\\

{\bf Fig.12.}  Double differential cross sections for the production of
$K^-$-mesons
 per proton, that interacted with Be nuclei, with $p_0=400$ GeV/c at
$p_{\perp}=0$ and $0.5$ GeV/c as functions of the $K^-$-meson momentum.\\

{\bf Fig.13.}  Spectra of $K^-$-mesons per proton, that interacted with Cu
nuclei,
with $p_0=24$ GeV/c at angles of 57 and 87 mrad. The experimental
 data: [14]. The solid and dashed curves, our calculation by (10)--(19),
 respectively, for angles of 57 and 87 mrad and $\sigma_a=850$ mb.\\

{\bf Fig.14.}  Dependence of the inclusive invariant cross sections for the
production of $K^-$-mesons in the collisions of protons with $p_0=100$ GeV/c
 with Cu and Pb nuclei on their momentum at $p_{\perp}=0.3$ GeV/c.
The experimental data: [13]. The dashed and solid curves, our calculation by
(10)--(19), respectively, for Cu and Pb nuclei.

\end{document}